\documentclass[letterpaper,titlepage,11pt]{article}
\usepackage{hyperref}

\usepackage{amssymb,amsmath,amsfonts}
\usepackage{mathrsfs}
\usepackage{epsfig}
\usepackage{verbatim}

\def\clock{{\count0=\time
           \divide\count0 60
           \ifnum\count0<10 0\fi\the\count0
           \multiply\count0 -60 \advance\count0 \time
           :\ifnum\count0<10 0\fi \the\count0
         }}

\newcommand{\timestamp}{{\small\vbox{\hbox{\tt\jobname.tex}
\hbox{\the\day/\the\month/\the\year, \clock}}}}

\newcommand{\CN}{\mathcal{N}}

\newcommand{\Z}{\mathbb{Z}}
\newcommand{\C}{\mathbb{C}}
\newcommand{\R}{\mathbb{R}}

\newcommand{\nn}{\nonumber}
\newcommand{\lspa}{\ \ ,\ \ \ \ }
\newcommand{\spa}{\,, \ \ }

\newcommand{\vecto}[2]{\left( \begin{array}{c} #1 \\ #2 \end{array} \right) }
\newcommand{\matrto}[4]{\left( \begin{array}{cc} #1 & #2 \\ #3 & #4 \end{array} \right) }

\newcommand{\ads}{\mbox{AdS}}
\newcommand{\gym}{g_{\rm YM}}

\newcommand{\trasp}[1]{\left(#1\right)^T}
\newcommand{\sums}[2]{\sum_{#1}^{#2}}
\newcommand{\etagamma}{\sum_{k=1}^{4}\eta_{k}\gamma^{2k-1,2k}}

\def\neqf{{\cal N}=4}

\def\gym{g_{YM}}

\def\a'{\alpha'}

\def\susyc{supersymmetric}

\def\bg{background}

\def\vp{\varphi}

\def\CP{\mathbb{C}{\rm P}}
\def\adscp{${\rm AdS}_4 \times \CP^3$}

\numberwithin{equation}{section}

\begin{document}

\begin{titlepage}
\rightline{\vbox{\small\hbox{\tt NORDITA-2009-82} }}

\ \vspace{2.2cm} \

\centerline{\LARGE  \bf New Penrose Limits and AdS/CFT}

\vskip 1.8cm

\centerline{\large {\bf Gianluca Grignani$\,^{1}$, Troels Harmark$\,^{2}$, Andrea Marini$\,^{1}$ and {\bf Marta Orselli$\,^{3}$} }}

\vskip 0.5cm

\begin{center}
\sl $^1$ Dipartimento di Fisica, Universit\`a di Perugia,\\
I.N.F.N. Sezione di Perugia,\\
Via Pascoli, I-06123 Perugia, Italy\\
\vskip 0.4cm
\sl $^2$ NORDITA\\
Roslagstullsbacken 23,
SE-106 91 Stockholm,
Sweden \vskip 0.4cm
\sl $^3$ The Niels Bohr Institute  \\
\sl  Blegdamsvej 17, DK-2100 Copenhagen \O , Denmark \\
\end{center}
\vskip 0.5cm

\centerline{\small\tt grignani@pg.infn.it, harmark@nordita.org, andrea.marini@fisica.unipg.it, orselli@nbi.dk}

\vskip 1.5cm

\centerline{\bf Abstract} \vskip 0.2cm

\noindent
We find a new Penrose limit of $\mbox{AdS}_5 \times S^5$
giving the maximally supersymmetric pp-wave background with two
explicit space-like isometries. This is an important missing piece in studying the AdS/CFT correspondence in certain subsectors. In particular whereas the Penrose limit giving one space-like isometry is useful for the $SU(2)$ sector of $\CN=4$ SYM, this new Penrose limit is instead useful for studying the $SU(2|3)$ and $SU(1,2|3)$ sectors. In addition to the new Penrose limit of $\mbox{AdS}_5 \times S^5$ we also find a new Penrose limit of $\mbox{AdS}_4 \times \C P^3$.

\vskip 0.5cm \leftline{\timestamp}

\end{titlepage}

\pagestyle{plain} \setcounter{page}{1}

\tableofcontents


\section{Introduction}
\label{sec:intro}

AdS/CFT duality identifies $\CN=4$
superconformal Yang-Mills (SYM) theory with gauge group $SU(N)$ to type IIB superstring theory on the  $\mbox{AdS}_5\times S^5$ background~\cite{Maldacena:1997re,Gubser:1998bc,Witten:1998qj}.
The AdS/CFT correspondence relates gauge theory and string theory in
different regimes, thus, on the one hand, this makes it powerful as it
can be used to compute the strong coupling regime of either theory using the weak coupling limit of the other, on the other hand this makes it hard to test directly since it is not easy to find situations where approximate computations in both theories have an overlapping domain of validity.

In \cite{Berenstein:2002jq} a way out of this difficulty was presented by introducing a Penrose limit of the $\mbox{AdS}_5\times S^5$ background. Taking the Penrose limit one gets the maximally supersymmetric
pp-wave background~\cite{Blau:2001ne,Blau:2002dy} where
type IIB string theory can be quantized \cite{Metsaev:2001bj,Metsaev:2002re}. On the gauge theory side the Penrose limit corresponds to considering a certain sector of the operators. This enables one to compare directly the spectrum of operators in the planar limit of $\CN=4$ SYM to the energy spectrum of quantum strings on the pp-wave.
In \cite{Bertolini:2002nr} an alternative Penrose limit of $\mbox{AdS}_5\times S^5$ was found also giving the maximally supersymmetric background but in a coordinate system with an explicit space-like isometry \cite{Michelson:2002wa,Bertolini:2002nr}. As explained in \cite{Harmark:2006ta} having this explicit isometry makes it particularly well-suited to study the $SU(2)$ sector of $\CN =4$ SYM.

Building on the Penrose limit of \cite{Berenstein:2002jq} many very interesting results in matching gauge theory and string theory were found in the case of the planar limit using the idea of integrability and the connection to
spin chains~\cite{Minahan:2002ve,Beisert:2003tq,Beisert:2003yb}\footnote{Early attempts in describing gauge theories in terms of spin chains can be found in~\cite{Berruto:1997jv,Berruto:1999ga}.} particularly by considering a near plane wave limit with curvature corrections to the pp-wave background \cite{Callan:2003xr,Callan:2004uv}. A high point of this is the development of the Asymptotic Bethe Ansatz describing the dimension of infinitely long operators for any 't Hooft coupling in the planar limit \cite{Staudacher:2004tk,Beisert:2005tm,Beisert:2006ez}. Going beyond the planar limit seems instead to be very difficult~\cite{Kristjansen:2002bb}.
New ideas are needed in order to further explore the AdS/CFT correspondence in the non-planar limit and its potential applications.

Recently another example
of an exact duality between $\CN = 6$ superconformal Chern-Simons theory (ABJM theory) and type IIA string theory on $\mbox{AdS}_4 \times CP^3$ have been found~\cite{Aharony:2008ug}.
Also here certain Penrose limits and near plane wave limits have been explored \cite{Nishioka:2008gz,Gaiotto:2008cg,Grignani:2008is,Astolfi:2008ji,Astolfi:2009qh}.

The difficulty of going beyond the planar limit, where integrability most likely is absent, makes it desirable to consider alternative approaches to match the spectrum of operators and string states. One of the cornerstones in comparing the operator spectrum to the string spectrum in a Penrose limit or near-plane wave limit is that in comparing the spectrum of operators one assumes that most of the operators of the gauge theory receive an infinitely large correction to the bare dimension in the large 't Hooft coupling limit $\lambda \rightarrow \infty$. This is of course a built in feature of the Asymptotic Bethe Ansatz for $\CN=4$ SYM. However, an alternative approach to this problem of taking the strong coupling limit of
$\CN=4$ SYM has been proposed
in~\cite{Harmark:2006di,Harmark:2006ta,Harmark:2006ie,Harmark:2007et,
Harmark:2007px, Harmark:2008gm} where a regime of AdS/CFT was found
in which both gauge theory and string theory are reliable and the
correspondence can be tested in a precise way.

Applying the approach of \cite{Harmark:2006di,Harmark:2006ta,Harmark:2006ie,Harmark:2007et,
Harmark:2007px,Harmark:2008gm}%
\footnote{See also \cite{Astolfi:2008yw} for work on the winding state with the space-like isometry compactified.} to match the spectrum of operators and string states in the $SU(2)$ sector uses in an essential way the alternative Penrose limit of \cite{Bertolini:2002nr} where the maximally supersymmetric pp-wave has an explicit isometry. This is because for this pp-wave background the string states having an energy just above the vacuum energy are the states dual to the operators in the $SU(2)$ sector of $\CN=4$ SYM.

However, as shown in \cite{Harmark:2007px} there are several other sectors of $\CN=4$ SYM that one can explore as well, and these sectors are crucial for approaching non-perturbative physics of type IIB string theory in $\mbox{AdS}_5\times S^5$, such as D-branes and black holes. This means that there should be additional Penrose limits of $\mbox{AdS}_5\times S^5$ in addition to the ones of \cite{Blau:2002dy,Berenstein:2002jq,Bertolini:2002nr}.

In this paper we address these issues by deriving a new Penrose limit of $\mbox{AdS}_5 \times S^5$ which leads to a new pp-wave background with two explicit space-like isometries. As for the two previously found Penrose limits \cite{Blau:2002dy,Berenstein:2002jq,Bertolini:2002nr} this leads to a pp-wave background where type IIB string
theory can be quantized and the spectrum can be matched to the spectrum of operators of $\CN=4$ SYM.
Our analysis completes the study of all possible pp-wave backgrounds which can be obtained as Penrose limits of the $\mbox{AdS}_5 \times S^5$  geometry.
It also represents a further step in the investigation of the matching of strongly coupled gauge theory and string theory in certain sectors which are relevant for describing non-perturbative physics of type IIB string theory on $\mbox{AdS}_5\times S^5$. In particular, the new Penrose limit is relevant for studying the $SU(1,2|3)$ sector, which is the maximally possible subsector of $\CN=4$ SYM \cite{Harmark:2007px}.

In addition to the new Penrose limit of $\mbox{AdS}_5\times S^5$ we also explore Penrose limits of $\mbox{AdS}_4 \times \C P^3$. Here two different classes of Penrose limits have been found, one in which there are no explicit space-like isometries \cite{Nishioka:2008gz,Gaiotto:2008cg} and another in which there are two explicit space-like isometries \cite{Grignani:2008is,Astolfi:2009qh} which makes it suitable for studying the $SU(2)\times SU(2)$ sector of ABJM theory. We find in this paper a new Penrose limit of the $\mbox{AdS}_4 \times \C P^3$ background giving a pp-wave background with one explicit space-like isometry.

The new Penrose limit of $\mbox{AdS}_5\times S^5$ found in this paper is also relevant for studying the finite temperature behavior of AdS/CFT. It is conjectured that the confinement/deconfinement transition temperature of planar $\mathcal{N}=4$ SYM on $R\times S^3$ is dual to the Hagedorn temperature of type IIB string theory on $\mbox{AdS}_5 \times S^5$ \cite{Witten:1998zw,Sundborg:1999ue,Polyakov:2001af,Aharony:2003sx}.
Using the Penrose limit \cite{Bertolini:2002nr} this was shown quantitatively to be true \cite{Harmark:2006ta} by matching the confiment/deconfinement temperature of planar $\mathcal{N}=4$ SYM on $R\times S^3$ in a limit with R-charge chemical potentials to the Hagedorn temperature of type IIB string on the pp-wave background of \cite{Bertolini:2002nr}\footnote{For related computations of the Hagedorn temperature in the presence of background
fields see for example Refs.~\cite{Deo:1989bv}.}.
We furthermore expect that our results
could help in understanding more generally the behavior of string
theory above the Hagedorn temperature and to study the connection
between gauge theory and black holes in $\mbox{AdS}_5 \times S^5$
\cite{Grignani:2009ua}\footnote{See also~\cite{Harmark:1999xt} for
a related study of black holes with R-charged chemical potentials.}.

Interesting related work in other less supersymmetric gauge theories
can be found in Refs.~\cite{Grignani:2007xz, Larsen:2007bm,Hamilton:2007he}.

The paper is organized as follows. In Section \ref{sec:stringtheory}
we first review the Penrose limit of string theory that lead to
pp-wave backgrounds with zero and one spatial isometry. Then, we
find a new Penrose limit giving rise to a pp-wave background with
two space-like isometries in which string theory can be quantized.
In Section \ref{sec:stringrotspectra} we obtain a general form for a
pp-wave metric that reproduces all the pp-wave backgrounds analyzed
in the previous section. We moreover show that string theory can be
directly quantized on this background which we dub ``{\em rotated
pp-wave background} " and we compute the spectrum. In
Section~\ref{sec:decsectors} we show that, after taking an
appropriate limit, the spectrum of type IIB string theory on the
rotated pp-wave background can be exactly matched to the spectrum of
the dual gauge theory operators in certain decoupled sectors of $\CN=4$ SYM. Finally, in Section~\ref{sec:ads4} we find a new Penrose limit of the $\ads_4\times \C P^3$ background of type IIA supergravity with one explicit space-like isometry.

\section{Penrose limits and pp-waves with explicit isometries}
\label{sec:stringtheory}

In this section we derive a Penrose limit of $\mbox{AdS}_5 \times S^5$
which results in a new pp-wave background with two space-like
isometries. We then show how to obtain a general pp-wave
background which, for appropriate choices of the parameters of the background, reproduces all the known pp-wave backgrounds which are obtained through a Penrose limit procedure on $\mbox{AdS}_5 \times S^5$. We begin the section by writing down a slightly generalized version of the previously found Penrose limits of $\mbox{AdS}_5 \times S^5$ with zero and one explicit space-like isometries \cite{Blau:2002dy,Berenstein:2002jq,Bertolini:2002nr}.

In $\mbox{AdS}_5 \times S^5$, the Penrose limit consists in considering a
particle in the center of $\mbox{AdS}_5 $ that is moving very rapidly on a
geodesic of $S^5$. This means that the angular momentum along the
direction in which the particle is moving is very large ($J \to
\infty$). Then by taking the limit $R \to \infty$, where $R$ is the radius  of $\mbox{AdS}_5$ and $S^5$, but such that the ratio
$J/R^2$ remains fixed, the geometry of $\mbox{AdS}_5 \times S^5$ reduces to
a plane-wave geometry.

An important point to emphasize is that one can choose any
light-like geodesic of $\mbox{AdS}_5 \times S^5$ for implementing the
procedure. While the pp-wave background always corresponds to the
maximally supersymmetric pp-wave background of type IIB supergravity
\cite{Blau:2001ne}, different choices of light-like geodesics can
give this background in different coordinate systems
\cite{Bertolini:2002nr}. Naively this should not matter, however,
the different coordinate systems can correspond to different choices
of lightcone time on the pp-wave background. And this corresponds
moreover to different dictionaries between the physical quantities
of the $\mbox{AdS}_5\times S^5$ background and of the maximally
supersymmetric pp-wave background. Therefore, the different
coordinate systems for the pp-wave background are connected to the
fact that the different Penrose limits that we consider correspond
to zooming in to different regimes of type IIB string theory on
$\mbox{AdS}_5\times S^5$. This in turns corresponds to zooming in to different regimes of $\CN=4$ SYM. Furthermore, as we discuss in section
\ref{sec:decsectors}, the different Penrose limits correspond to different decoupling limits of
$\CN=4$ SYM on $\R\times S^3$.

In the literature the ``canonical'' coordinate system used for the
maximally supersymmetric pp-wave background is that of
\cite{Blau:2001ne,Blau:2002dy,Berenstein:2002jq} which we here dub
the  {\em BMN pp-wave background}. This coordinate system is such
that the quadratic potential terms for the transverse directions are
massive for all eight transverse directions. Another coordinate
system was introduced in~\cite{Michelson:2002wa,Bertolini:2002nr} and we will refer
to it as the {\em one flat direction pp-wave background} due to the
presence of a space-like isometry in the pp-wave metric and since in
this case the quadratic terms for the transverse directions have a
massless direction.

Here we find a new pp-wave background corresponding to a new
coordinate system for the maximally supersymmetric pp-wave of type
IIB supergravity. This new background is  again obtained as a
Penrose limit of $\mbox{AdS}_5 \times S^5$ with an appropriate choice of
light-cone coordinates. The new pp-wave background differs from the
other two because of the presence of two spacial isometries in the
metric, namely two flat directions, corresponding to two massless
directions in the potential terms for the transverse directions.
Hence we call it the {\em two flat directions pp-wave background}.

This new pp-wave background is important in the context of the
AdS/CFT correspondence. In fact, as shown explicitly in Section \ref{sec:stringrotspectra}, string theory can be quantized on this background. Moreover, as discussed in Section \ref{sec:decsectors},
after taking a certain limit on the spectrum of type IIB
string theory in this new background, we can complete the
matching between the spectrum of anomalous dimensions of gauge theory operators in certain sectors of $\neqf$ SYM theory and the spectrum of the dual string theory states.

We show below in Section \ref{sec:stringrotspectra} that all the pp-wave \bg s achievable through the
Penrose limit are connected by a time-dependent coordinate
transformation. This proves that mathematically they are all
equivalent. The same is not true from the physical point of view,
since the transformation involves time. Thus what changes from a
\bg\ to another is what we call time, and consequently what we call
Hamiltonian. Therefore the physics is different when we consider the
theory on different pp-wave backgrounds.

It is also interesting to notice which regimes of $\CN=4$ SYM the different Penrose limits correspond to. We give these regimes for each of the three different limits below. To consider this, we record the following dictionary between strings on $\mbox{AdS}_5\times S^5$ and $\CN=4$ SYM on $\R \times S^3$. We have
\begin{equation}
\frac{R^4}{l_s^4} = 4 \pi^2 \lambda \spa g_s = \frac{\pi \lambda}{N}
\end{equation}
where $R$ is the radius of $\mbox{AdS}_5$ and $S^5$, $g_s$ and $l_s$ are the string coupling and string length, respectively, and $\lambda = \gym^2 N/(4\pi^2)$ is the 't Hooft coupling of $SU(N)$ $\CN=4$ SYM.\footnote{The $4\pi^2$ factor is included in the 't Hooft
coupling for our convenience.} The energy $E$ of type IIB string states on $\mbox{AdS}_5\times S^5$ is identified with the energy $E$ of the dual $\CN=4$ SYM states on $\R \times S^3$, or equivalently, with the scaling dimension of the dual operators of $\CN=4$ SYM on $\R^4$. Similarly the angular momenta $J_{1,2,3}$ on $S^5$ for string states are identified with the three R-charges $J_{1,2,3}$ for states/operators of $\CN=4$ SYM. Moreover the angular momenta $S_{1,2}$ for strings on $\mbox{AdS}_5$ are identified with the Cartan generators for the $SO(4)$ symmetry of the $S^3$ for the dual $\CN=4$ SYM states on $\R \times S^3$, or equivalently, the $SO(4)$ symmetry of the $\R^4$ for the dual operators of $\CN=4$ SYM on $\R^4$.

The string theory that we are interested in is type IIB string theory on
$\mbox{AdS}_5 \times S^5$. The metric for this background is given by
\begin{equation}
\label{adsmet}
ds^2 = R^2 \left[ - \cosh^2 \rho dt^2 + d\rho^2 +
\sinh^2 \rho d{\Omega'_3}^2 + d\theta^2 + \sin^2 \theta d\alpha^2
+ \cos^2 \theta d\Omega_3^2 \right]\, ,
\end{equation}
with the five-form Ramond-Ramond field strength
\begin{equation}
\label{adsF5}
F_{(5)} = 2 R^4 ( \cosh \rho \sinh^3 \rho dt d\rho d\Omega_3' +
 \sin \theta \cos^3 \theta d\theta d\alpha d\Omega_3 )\, .
\end{equation}
We parameterize the two three-spheres as
\begin{align}
\label{3sph}
d\Omega_3^2 &= d\psi^2 + \sin^2 \psi d\phi^2 + \cos^2 \psi d\chi^2\, , \\
\label{3sphAdS}
d\Omega_3'^2 &= d\beta^2 + \sin^2 \beta d\gamma^2 + \cos^2 \beta d\xi^2\, .
\end{align}
The three angular momenta on the five sphere $S^5$ are defined
as
\begin{align}
\label{eq:JJJ}
J_1= -i\partial_\chi\, , \quad J_2= -i\partial_\phi\, , \quad J_3= -i\partial_\alpha\, ,
\end{align}
and the two angular momenta on the $S^3$ inside $\mbox{AdS}_5$ are defined
as
\begin{align}
\label{eq:SS}
S_1 = -i \partial_\gamma\, , \qquad S_2=-i\partial_\xi \, .
\end{align}
We moreover define the quantity $J\equiv J_1 + \eta_1 J_2 + \eta_2 J_3 + \eta_3
S_1 + \eta_4 S_2$, where $\eta_1$, $\eta_2$, $\eta_3$, $\eta_4$ are
some parameters that characterize the background. We will show that they play an important role in Section \ref{sec:decsectors} where we compare the results we obtain on the string theory side with previous computations done in the dual gauge theory.

\subsection{The ``no flat direction'' Penrose limit}

In order to derive the new Penrose limit, we first review the
Penrose limit giving rise to the {\em BMN pp-wave \bg}. We introduce
new coordinates $\varphi_0,...,\varphi_4$ defined by
\begin{align}
\label{eq:noflatphi}
\chi &= \varphi_0, \quad
\phi = \eta_1 \varphi_0 +  \varphi_1\, , \quad
\alpha = \eta_2 \varphi_0 +  \varphi_2\, , \quad
\gamma = \eta_3 \varphi_0 +  \varphi_3\, , \quad
\xi = \eta_4 \varphi_0 +  \varphi_4\,,
\end{align}
and we define the light-cone coordinates as
\begin{align}
z^- = \frac{1}{2} \mu R^2 (t-\varphi_0)\, , \quad
z^+ = \frac{1}{2\mu} (t+\varphi_0)\, .
\label{lcc}
\end{align}
By defining $r_1,...,r_4$ such that
\begin{align}
r_1= R \psi\, , \quad
r_2 = R \theta\, ,\quad
r_3 = R \rho \sin\beta\, ,\quad
r_4= R \rho \cos\beta\, .
\end{align}
we can parametrize the eight $z^i$ coordinates in the following way
\begin{align}
\label{coordinates}
z^1+iz^2 = r_1e^{i\varphi_1}\, , \quad z^3+iz^4 = r_2e^{i\varphi_2}\, , \cr
z^5+iz^6 = r_3e^{i\varphi_3}\, , \quad z^7+iz^8 = r_4e^{i\varphi_4}\, .
\end{align}

Writing the background \eqref{adsmet}--\eqref{adsF5} in terms of the
coordinate $z^\pm$ and $z^i$ and taking the Penrose limit by sending
$R\to\infty$ while keeping $z^\pm$ and $z^i$ fixed, we obtain the
following metric
\begin{equation}\label{eq:dsnoflat}
    \begin{split}
        ds^2=&-4dz^+dz^- + dz^i dz^i - \mu^2 \sum_{k=1}^{4}
        \left(1-\eta_{k}^{2}\right)\left[\left(z^{2k-1}\right)^2+\left(z^{2k}\right)^2\right]\left(dz^+\right)^2 \\
        &+ 2\mu \sum_{k=1}^{4}\eta_k \left[z^{2k-1}dz^{2k}- z^{2k}dz^{2k-1}\right]dz^+.
    \end{split}
\end{equation}
and five-form field strength
\begin{align}
\label{eq:F5z}
F_{(5)} = 2 \mu \,dz^+ \left(dz^1 dz^2 dz^3 dz^4 + dz^5 dz^6 dz^7 dz^8 \right)\, .
\end{align}
We see that by setting the parameters $\eta_k$'s all to zero, we
precisely recover the pp-wave background derived in~
\cite{Blau:2002mw,Berenstein:2002jq}. In this sense, the background
\eqref{eq:dsnoflat}--\eqref{eq:F5z} is a generalization of it. Type
IIB string theory can be quantized on this background and the
light-cone Hamiltonian that one obtains is
\begin{align}
H_\textrm{lc} \sim E-J_1, \qquad  p^+ \sim \frac{E+J_1}{R^2}\, .
\end{align}
From the condition that $H_\textrm{lc}$  and $p^+$ should stay finite in the limit, we get
that $J_1=-i\partial_{\varphi_0}$ must be large. On the other hand
 since $\varphi_1, ..., \varphi_4$ are all fixed in the
limit $R\to \infty$, we deduce from \eqref{eq:JJJ}, \eqref{eq:SS} and \eqref{eq:noflatphi}
that $J_2$, $J_3$, $S_1$ and $S_2$ are also fixed.

We see from the above that the ``no flat direction'' Penrose limit corresponds to the following regime of type IIB string theory on $\mbox{AdS}_5\times S^5$
\begin{equation}
R \rightarrow \infty \ \mbox{with}\ E-J_1\ \mbox{fixed} , \quad \frac{E+J_1}{R^2}\ \mbox{fixed}, \quad \frac{J_1}{R^2} \ \mbox{fixed}, \quad g_s,l_s \ \mbox{fixed}
\end{equation}
Translating this into $\CN=4$ SYM language, it corresponds to the regime
\begin{equation}
N \rightarrow \infty \ \mbox{with}\ E-J_1\ \mbox{fixed} , \quad \frac{E+J_1}{\sqrt{N}}\ \mbox{fixed}, \quad \frac{J_1}{\sqrt{N}} \ \mbox{fixed}, \quad \gym^2 \ \mbox{fixed}
\end{equation}

\subsection{The ``one flat direction'' Penrose limit}

Now we repeat an analogous procedure and show that, by a different
choice of light-cone coordinates, we obtain a generalization of the
pp-wave background derived in \cite{Bertolini:2002nr}. We  define
the coordinates $\varphi_0,...,\varphi_4$ in the following way
\begin{align}
\label{eq:oneflatphi}
\chi = \varphi_0 -\varphi_1\, , \quad
\phi = \varphi_0 +  \varphi_1\, , \quad
\alpha = \eta_2 \varphi_0 +  \varphi_2\, , \quad
\gamma = \eta_3\varphi_0 + \varphi_3\, , \quad
\xi = \eta_4\varphi_0 + \varphi_4\, ,
\end{align}
with the light-cone variables still given by eq.n \eqref{lcc}.

We moreover define $z^1$ and $z^2$ as
\begin{align}
z^1 = R\varphi_1\, , \quad z^2=R\left(\frac{\pi}{4}-\psi\right)\, ,
\end{align}
while $z^3,...,z^8$ are defined as before (see Eq.\eqref{coordinates}) and
\begin{align}
r_2 = R \theta\, , \quad
r_3 = R \rho \sin\beta\, ,\quad
r_4= R \rho \cos\beta\, ,
\end{align}
\begin{align}
z^3+iz^4 =r_2 e^{i\varphi_2}\, , \quad
z_5+iz_6 = r_3e^{i\varphi_3}\, , \quad z_7+iz_8 = r_4e^{i\varphi_4}\, .
\end{align}
The Penrose limit is then the limit $R\to\infty$ keeping $z^\pm,z^i$ fixed.
Plugging the coordinates $z^\pm, z^i$ into the background
\eqref{adsmet}--\eqref{adsF5} and taking the limit described above
the metric becomes
\begin{equation}\label{eq:dsoneflat}
    \begin{split}
        ds^2=&-4dz^+dz^- + dz^i dz^i - \mu^2 \sum_{k=2}^{4}
        \left(1-\eta_{k}^{2}\right)\left[\left(z^{2k-1}\right)^2+\left(z^{2k}\right)^2\right]\left(dz^+\right)^2 \\
        &+ 2\mu \sum_{k=2}^{4}\eta_k \left[z^{2k-1}dz^{2k}- z^{2k}dz^{2k-1}\right]dz^+ -4 \mu z^2 dz^+ dz^1.
    \end{split}
\end{equation}
with the five-form given by~\eqref{eq:F5z}.

From \eqref{eq:dsoneflat} we see that $z^1$ is an explicit isometry
of the above pp-wave background and therefore we call this
background {\em one flat direction pp-wave background}.

As before we have that $\varphi_2,\varphi_3,\varphi_4$ are fixed in the Penrose
limit which, using \eqref{eq:oneflatphi}, means that $J_3$, $S_1$ and $S_2$ are
fixed.  But now the condition that $H_\textrm{lc}$, $p^+$ and $p^1$
have to remain finite in the limit tells us that the quantities
\begin{equation}
E-J_1-J_2 , \quad \frac{E+J_1+J_2}{R^2}, \quad \frac{J_1+J_2}{R^2} , \quad \frac{J_1-J_2}{R} , \quad g_s,l_s
\end{equation}
are all fixed when $R \to \infty$. This is the regime corresponding to the ``one flat direction'' Penrose limit of type IIB string theory on $\mbox{AdS}_5\times S^5$, as found in \cite{Bertolini:2002nr}.
Translating this into $\CN=4$ SYM language, it corresponds to the regime where \cite{Bertolini:2002nr}
\begin{equation}
E-J_1-J_2 , \quad \frac{E+J_1+J_2}{\sqrt{N}}, \quad \frac{J_1+J_2}{\sqrt{N}} , \quad \frac{J_1-J_2}{N^{1/4}} , \quad \gym^2
\end{equation}
are fixed for $N \to \infty$.

\subsection{The ``two flat directions'' Penrose limit}

We finally consider the Penrose limit that leads to a new pp-wave \bg\ with two flat directions.
The variables $\varphi_0,$ $\varphi_1,$ $\varphi_2,$  $\varphi_3,$
$\varphi_4$ are now defined as
\begin{gather}\label{phi2fd}
        \chi  = \varphi_0 - \sqrt{2}\varphi_1 - \varphi_2 \,  , \qquad
        \phi  = \varphi_0 + \sqrt{2}\varphi_1 - \varphi_2\,  , \qquad
        \alpha  = \varphi_0 + \varphi_2 \,  ,\nn \\[2mm]
        \gamma  = \eta_3 \varphi_0 + \varphi_3  \, , \qquad
        \xi  =  \eta_4 \varphi_0 + \varphi_4 \, ,
\end{gather}
whereas the light-cone coordinate are as usual given by \eqref{lcc}.

The coordinates $z^1$, $z^2$, $z^3$ and $z^4$ are defined as
\begin{equation}
    \begin{array}{lcl}
        z^1  = R \varphi_1 \,  , & \phantom{qquad} & z^2  =
        \displaystyle{ \frac{R}{\sqrt{2}}} \left(\displaystyle{\frac{ \pi}{4}-\psi}\right) \,  , \\[4mm]
        z^3 = R \varphi_2 \,  , &  & z^4  = R \left(\displaystyle{\frac{ \pi}{4}}-\theta \right) \,  .
    \end{array}
\end{equation}
while $z^5$, $z^6$, $z^7$, $z^8$ are again given by Eq.\eqref{coordinates}. More explicitly we have
\begin{align}
    r_3  = R \rho \sin \beta \, , \qquad     r_4  = R \rho \cos \beta\,  ,
\end{align}
\begin{align}
    z^5 + i z^6  = r_3 \displaystyle{ e^{i \varphi_3}} \, , \qquad z^7  + i z^8 = r_4 \displaystyle{ e^{i \varphi_4}}\,  .
\end{align}
Substituting the new coordinates in the background \eqref{adsmet}--\eqref{adsF5}
and taking the Penrose limit we get the following pp-wave metric
\begin{equation}\label{eq:dstwoflat}
    \begin{split}
        ds^2&=-4dz^+dz^- + dz^i dz^i - \mu^2 \sum_{k=3,4}
        \left(1-\eta_{k}^{2}\right)\left[\left(z^{2k-1}\right)^2+\left(z^{2k}\right)^2\right]\left(dz^+\right)^2 \\
        &+ 2\mu \sum_{k=3,4}\eta_k\left[ z^{2k-1}dz^{2k}-  z^{2k}dz^{2k-1}\right]dz^+
        - 4\mu\left(z^2 dz^1 + z^4 dz^3\right)dz^+.
    \end{split}
\end{equation}
and the five-form is defined in \eqref{eq:F5z}. This is a new
pp-wave background and it has two explicit isometries, $z^1$ and
$z^3$ We will therefore refer to it as {\em two flat directions
pp-wave background}.

In this case $\varphi_3,\varphi_4$ are fixed, thus, keeping in mind
\eqref{phi2fd}, we have that also the angular momenta $S_1$ and $S_2$ are
fixed. In a similar fashion as before if we compute $H_\textrm{lc}$, $p^+$, $p^1$
and $p^3$ and request that they should stay finite in the Penrose limit
we get that the quantities
\begin{equation}
E-J_1-J_2-J_3 , \quad \frac{E+J_1+J_2+J_3}{R^2},
\quad
\frac{J_1+J_2+J_3}{R^2}, \quad \frac{J_1 - J_2}{R} ,\quad  \frac{J_3 -J_1 - J_2}{R}, \quad g_s,l_s
\end{equation}
are fixed as $R$ goes to infinity. This is the regime corresponding to the ``two flat directions'' Penrose limit of type IIB string theory on $\mbox{AdS}_5\times S^5$.
Translating this into $\CN=4$ SYM it corresponds to the regime where
\begin{equation}
E-J_1-J_2-J_3 , \quad \frac{E+J_1+J_2+J_3}{\sqrt{N}}, \quad
\frac{J_1+J_2+J_3}{\sqrt{N}}, \quad \frac{J_1 - J_2}{N^{1/4}} ,\quad  \frac{J_3 -J_1 - J_2}{N^{1/4}} , \quad \gym^2
\end{equation}
are fixed for $N \rightarrow \infty$. Here $J_1-J_2$ and $J_3-J_1-J_2$ correspond to the two momenta for the two space-like isometries of the {\em two flat directions pp-wave background} \eqref{eq:dstwoflat}.

Type IIB string theory on the pp-wave backgrounds \eqref{eq:dsnoflat}, \eqref{eq:dsoneflat}
\eqref{eq:dstwoflat} (with five-form field strength given by \eqref{eq:F5z})
can be easily quantized. The spectra in all these three cases
are worked out in the next section.


\section{String theory spectrum on a rotated pp-wave background}
\label{sec:stringrotspectra}

In this section we obtain a pp-wave metric, which depends on
parameters introduced through a coordinate transformation on the
maximally \susyc\ \bg\ of \cite{Blau:2001ne}. For this reason, in
practice, this metric describes an infinite set of pp-wave \bg s (one
for each point of the parameter space). We refer to them as to
\emph{rotated pp-wave backgrounds}.

Note that the backgrounds obtained in this way do not necessarily have any
specific meaning in an AdS/CFT context. They will only have a meaning in the AdS/CFT context if we derive them from a Penrose limit of $\mbox{AdS}_5 \times S^5$. Despite this, the procedure that we are going to show results to be very
useful because allows to obtain a general formula that contains all
the physically interesting pp-wave \bg s. In fact we will show that
by appropriately choosing the values of the parameters of the
background, this general formula describes exactly the \bg s studied
in the previous section which are indeed obtained by taking Penrose
limits of the $\mbox{AdS}_5 \times S^5$ \bg.

We can then proceed in finding the spectra on these generic rotated
\bg s. An important result is that, by taking an appropriate limit
on these spectra, we will show that one can reproduce the spectra
found in \cite{Harmark:2007px} for the nine decoupled sectors of
$\neqf$ SYM which contain scalars.

\subsection{Coordinate transformation}

We start from the simplest pp-wave background metric
without flat directions
\begin{equation}\label{BMNmetric}
ds^2=-4dx^+dx^- - \mu^2 x^ix^i\left(dx^+\right)^2+dx^idx^i\, ,
\end{equation}
where $i=1,2,\dots,8$ and five-form field strength
\begin{equation}\label{fff}
    F_{(5)}=2\mu dx^{+}\left(dx^{1}dx^{2}dx^{3}dx^{4}+dx^{5}dx^{6}dx^{7}dx^{8}\right)\, .
\end{equation}
We consider the following coordinate transformation
\begin{equation}\label{transfrot}
\begin{split}
x^- =z^- &+\frac{\mu}{2}\left(C_1 z^1z^2 + C_2z^3z^4 + C_3z^5z^6 + C_4z^7z^8\right)\, , \\[2mm]
\left( \begin{array}{c}
 x^{2k-1} \\[2mm]
 x^{2k}
\end{array} \right) &=
\left( \begin{array}{cc}
 \cos(\eta_k \mu z^+) & -\sin(\eta_k \mu z^+) \\[2mm]
 \sin(\eta_k \mu z^+) & \cos(\eta_k \mu z^+)
\end{array} \right)
\left( \begin{array}{c}
 z^{2k-1} \\[2mm]
 z^{2k}
\end{array} \right)\, ,
\end{split}
\end{equation}
where $C_{k}$ and $\eta_{k}$, $k=1, 2, 3, 4$, are parameters.

Note that the transformations for the transverse coordinates are rotations whose
angles depend on the $\eta_k$ parameters, hence the name ``{\em rotated pp-wave \bg}''.

The metric \eqref{BMNmetric} then becomes
\begin{equation}\label{rotmetric}
    \begin{split}
        ds^2=&-4dz^+dz^- + dz^i dz^i - \mu^2 \sum_{k=1}^{4}     \left(1-\eta_{k}^{2}\right)\left[\left(z^{2k-1}\right)^2+\left(z^{2k}\right)^2\right]\left(dz^+\right)^2 \\
        &- 2\mu \sum_{k=1}^{4}\left[(C_k-\eta_k)z^{2k-1}dz^{2k}+(C_k+\eta_k)z^{2k}dz^{2k-1}\right]dz^+\, ,
    \end{split}
\end{equation}
while the five-form field strength \eqref{fff} is invariant under
the coordinate transformation \eqref{transfrot}. It is
straightforward to check that the metric \eqref{rotmetric} contains
all the \bg s obtained in Section \ref{sec:stringtheory}. In fact,
for various values of the $C_k$ and $\eta_k$ parameters, we have
the following possibilities
\begin{center}
        \begin{tabular}{lcl}
            $C_1=C_2=C_3=C_4=0$ & $\Rightarrow$ & no flat direction; \\
            $C_1=\eta_1=1$ and $C_2=C_3=C_4=0$ & $\Rightarrow$ & one flat direction; \\
            $C_1=\eta_1=C_2=\eta_2=1$ and $C_3=C_4=0$ & $\Rightarrow$ & two flat directions.
        \end{tabular}
\end{center}

String theory can be quantized on the general background~\eqref{rotmetric} and we now
proceed in finding the superstring spectrum.

\subsection{Bosonic sector}

We work in the light-cone gauge $z^+ = p^+ \tau$ with $l_s=1$. The light-cone Lagrangian density of the bosonic $\sigma$-model is given by
\begin{equation}\label{boslagr}
\begin{split}
    \mathscr{L}_{lc}^{B}= &- \frac{1}{4\pi  p^+}\left(\partial^{\alpha}z^i\partial_{\alpha}z^i+
     f^2 \sum_{k=1}^{4}\left(1-\eta_{k}^{2}\right)\left[\left(z^{2k-1}\right)^2+\left(z^{2k}\right)^2\right] \right.  \\
    &+\left. 2f \sum_{k=1}^{4}\left[(C_k-\eta_k)z^{2k-1}\dot{z}^{2k}+(C_k+\eta_k)z^{2k}\dot{z}^{2k-1}\right]\right)\, ,
\end{split}
\end{equation}
where we have defined $f = \mu  p^+$. The conjugate momenta are computed to be
\begin{equation}
\Pi_{2k-1} = \frac{\dot{z}^{2k-1}-f\left(C_k + \eta_k \right) z^{2k}}{2\pi   }\, ,~~~~~
\Pi_{2k} = \frac{\dot{z}^{2k}-f\left(C_k - \eta_k \right) z^{2k-1}}{2\pi   }\, ,
\end{equation}
and the bosonic light-cone Hamiltonian is given by
\begin{equation}
 H_{lc}^{B}= \frac{1}{4\pi  p^+}\int_{0}^{2\pi}d\sigma \Bigg[ \dot{z}^i \dot{z}^i+ (z^i)'(z^i)'
  +f^2 \sum_{k=1}^{4}\left(1-\eta_{k}^{2}\right)\left[\left(z^{2k-1}\right)^2+\left(z^{2k}\right)^2\right]\Bigg]\, .
\end{equation}
In order to solve the equations of motion
\begin{subequations}
\begin{align}
&\partial^{\alpha}\partial_{\alpha}z^{2k-1}+2f\eta_k \dot{z}^{2k} - f^2 \left(1-\eta_{k}^{2}\right) z^{2k-1}=0\label{moteq1}\, ,\\
&\partial^{\alpha}\partial_{\alpha}z^{2k}-2f\eta_k \dot{z}^{2k-1} - f^2 \left(1-\eta_{k}^{2}\right) z^{2k}=0\label{moteq2}\, ,
\end{align}
\end{subequations}
it is useful to introduce four complex fields
\begin{equation}
    X^k = z^{2k-1}+ iz^{2k}\, ,
\end{equation}
in terms of which the above equations read
\begin{subequations}
\begin{align}
&\partial^{\alpha}\partial_{\alpha}X^{k}-2 i f\eta_k \dot{X}^{k} - f^2 \left(1-\eta_{k}^{2}\right) X^{k}=0\, ,\label{moteqd1}\\
&\partial^{\alpha}\partial_{\alpha}\bar{X}^{k}+2 i f\eta_k \dot{\bar{X}}^{k} - f^2 \left(1-\eta_{k}^{2}\right) \bar{X}^{k}=0\label{moteqd2}\, .
\end{align}
\end{subequations}
One can see that a solution of the form
\begin{equation}
    X^k=e^{-i f \eta_k \tau} Y^k
\end{equation}
solves \eqref{moteqd1} if $Y^k$ satisfy the equation
\begin{equation}
    \partial^{\alpha}\partial_{\alpha}Y^{k} -f^2 Y^k=0\, .
\end{equation}
Therefore for $Y^k$ and its conjugate $\bar{Y}^k$ we have the following mode expansions
\begin{subequations}\label{bosmodeex}
\begin{align}
Y^k&=i \sum_{n=-\infty}^{+\infty} \frac{1}{\sqrt{\omega_n}}\left(a_{n}^{k}e^{-i (\omega_n \tau -n\sigma)}- \left(\tilde{a}_{n}^{k}\right)^\dagger e^{i (\omega_n \tau -n\sigma)}\right)\, , \\
\bar{Y}^k&=i   \sum_{n=-\infty}^{+\infty} \frac{1}{\sqrt{\omega_n}}\left(\tilde{a}_{n}^{k}e^{-i (\omega_n \tau -n\sigma)}- \left(a_{n}^{k}\right)^\dagger e^{i (\omega_n \tau -n\sigma)}\right)\, .
\end{align}
\end{subequations}
The bosonic Hamiltonian now reads
\begin{equation}\label{Hcomplexfield}
     H_{lc}^{B}=  \frac{1}{4\pi p^+}\int_{0}^{2\pi}d\sigma \sum_{k=1}^{4}\left(\dot{\bar{X}}^k \dot{X}^k+ (\bar{X}^k)'(X^k)'
     +f^2 \left(1-\eta_{k}^{2}\right)\bar{X}^{k}X^{k}\right)\, .
\end{equation}
Then we quantize the theory imposing the canonical equal time commutation relations
\begin{equation}\label{etcr}
    \left[a_{n}^{k},a_{m}^{k'}\right]=0\, , \qquad \left[a_{n}^{k},(a_{m}^{k'})^{\dagger}\right]=\left[\tilde{a}_{n}^{k},(\tilde{a}_{m}^{k'})^{\dagger}\right]=\delta^{kk'}\delta_{nm}\, .
\end{equation}
We obtain the following bosonic spectrum in this background
\begin{equation}
\label{rotbosH}
\begin{split}
    H_{lc}^{B}=&  \frac{1}{ p^+}\sum_{n=-\infty}^{+\infty}  \sum_{k=1}^2
    \left[\left(\omega_n + \eta_k f\right) M_{n}^{(k)}+\left(\omega_n - \eta_k f\right) \tilde{M}_{n}^{(k)}\right. \\
    +&\left.\left(\omega_n + \eta_{(k+2)} f\right) N_{n}^{(k)}+\left(\omega_n - \eta_{(k+2)}  f\right) \tilde{N}_{n}^{(k)}\right]\, ,
\end{split}
\end{equation}
where
$\omega_n = \sqrt{n^2 + f^2}$ for all $n\in \mathbb{Z}$
and the number operators are defined as
\begin{equation}
M_{n}^{(k)}=a_{n}^{k\dagger}a_{n}^{k}\, , ~~ \tilde{M}_{n}^{(k)} =\tilde{a}_{n}^{k\dagger}\tilde{a}_{n}^{k}   \, ,~~N_{n}^{(k)}=a_{n}^{(k+2)\dagger}a_{n}^{(k+2)}\, , ~~\tilde{N}_{n}^{(k)} =\tilde{a}_{n}^{(k+2)\dagger}\tilde{a}_{n}^{(k+2)}
\end{equation}
for $k=1,2$.

\subsection{Fermionic sector}

We now work out the fermionic part of the spectrum.
The light-cone gauge and $\kappa$-symmetry gauge fixing condition are
\begin{equation}
    z^+ =  p^+ \tau, \qquad \Gamma^{+}\theta^A=0\,
\end{equation}
where $\theta^A$, with $A=1,2$, is a Majorana-Weyl spinor with $32$ components.
The Green-Schwarz fermionic light-cone action is then given by \cite{Metsaev:2002re}
\begin{equation}\label{GSaction}
    S_{lc}^{F}= \frac{i}{4\pi p^+}\int d\tau d\sigma \left[ \left(\eta^{\alpha\beta}\delta_{AB}-\epsilon^{\alpha\beta}\left(\sigma_{3}\right)_{AB}\right)\partial_{\alpha}z^+ \bar{\theta}^A \Gamma_+ \left(\mathcal{D}_{\beta}\theta\right)^B\right]\, ,
\end{equation}
with covariant derivative
\begin{equation}
    \mathcal{D}_{\alpha}=\partial_{\alpha}+\frac{1}{4}\partial_{\alpha}z^+ \left(\omega_{+\rho\sigma}\Gamma^{\rho \sigma}-\frac{1}{2\cdot 5!}F_{\lambda\nu\rho\sigma\kappa}\Gamma^{\lambda\nu\rho\sigma\kappa}i\sigma_2 \Gamma_+ \right)\, ,
\end{equation}
where $\sigma_{k}$'s are the Pauli matrices and $\omega_{a,b,c}$ are the spin connections.
The non-vanishing components of the five-form field strength are $F_{+1234}=F_{+5678}=2\mu$.

We can
write the action as
\begin{equation}\label{feract}
\begin{split}
    S_{lc}^{F}=& \frac{i}{2\pi  p^+ }\int d\tau d\sigma \Bigg\{\trasp{S^1}  \left[\partial_{+}-\frac{f}{2}\sum_{k=1}^{4}\eta_{k}\gamma^{2k-1,2k}\right]S^1\\
    +& \trasp{S^2}  \left[\partial_{-}-\frac{f}{2}\sum_{k=1}^{4}\eta_{k}\gamma^{2k-1,2k}\right]S^2 -2f \trasp{S^1} \Pi S^2\Bigg\}\, .
\end{split}
\end{equation}
where  $S^A$, $A=1,2$, is a eight component real spinor and we introduced the matrix $\Pi=\gamma^{1234}$, where $\gamma_i$ are $8\times 8$ Dirac matrices~\footnote{See Appendix \ref{AppendixA} for our conventions on the spinors and the representation of the Dirac matrices.}. Moreover, $\partial_{\pm}=\partial_{\tau}\pm\partial_{\sigma}$.
The equations of motion are
\begin{subequations}\label{eqmotferm}
\begin{align}
    &\left(\partial_{+}-\frac{f}{2}\sum_{k=1}^{4}\eta_{k}\gamma^{2k-1,2k}\right)S^{1}-f\Pi S^{2}=0\, ,\\
    &\left(\partial_{-}-\frac{f}{2}\sum_{k=1}^{4}\eta_{k}\gamma^{2k-1,2k}\right)S^{2}+f\Pi S^{1}=0\, .
\end{align}
\end{subequations}
It is useful to observe that a field of the form
\begin{equation}
    S^{A}=e^{\displaystyle \frac{f}{2}\sum_{k=1}^{4}\eta_{k}\gamma^{2k-1,2k}\tau}\Sigma^{A}
\end{equation}
satisfies the above equations if the fields $\Sigma^{A}$ obey the equations of motion of the
fermionic fields in the usual pp-wave background \cite{Metsaev:2001bj,Metsaev:2002re}:
\begin{equation}
\partial_{+}\Sigma^{1}-f\Pi \Sigma^{2}=0\, ,~~~~~~\partial_{-}\Sigma^{2}+f\Pi \Sigma^{1}=0\, ,
\end{equation}
whose solutions are
\begin{subequations}
\begin{align}
    &\Sigma^{1}=c_0\, e^{-i f \tau}S_0 - \sum_{n>0}c_n e^{-i \omega_{n}\tau}
    \left(S_n e^{i n \sigma}+\frac{\omega_{n}-n}{f} S_{-n}e^{-i n \sigma} \right)
    +\textrm{h.c. },\\
    &\Sigma^{2}=-c_0\, e^{-i f \tau}i\Pi S_0 - i \Pi\sum_{n>0}c_n e^{-i \omega_{n}\tau}
    \left(S_{-n} e^{-i n \sigma}-\frac{\omega_{n}-n}{f} S_{n}e^{i n \sigma} \right)+\textrm{h.c. },
\end{align}
\end{subequations}
where, for all values of $n$, $\omega_{n}=\sqrt{n^2+f^2}$, while $c_n =
\frac{1}{\sqrt{2}}[1+(\frac{\omega_{n}-n}{f})^{2}]^{-1/2}$.

The fermionic conjugate momenta can be computed from the action \eqref{feract}
\begin{equation}
    \lambda^{A}=\frac{i}{2\pi}S^{A}\, ,
\end{equation}
and the fermionic part of the Hamiltonian can be written in the form
\begin{equation}
    H_{lc}^{F}= \frac{i}{2\pi p^+ }\int^{2\pi}_{0}d\sigma
    \left(\trasp{S^1}\dot{S^1}+\trasp{S^2}\dot{S^2}\right)\,
\end{equation}
where we used the equations of motion \eqref{eqmotferm}.
Now we quantize the theory imposing the canonical equal time anticommutation relations
\begin{equation}
    \left\{S_{n}^{a},\left(S_{m}^{b}\right)^{\dagger}\right\}=\delta^{ab}\delta_{nm}\,
\end{equation}
and  the fermionic Hamiltonian reads
\begin{equation}
    H_{lc}^{F}=\frac{1}{ p^+ }\sums{n=-\infty}{+\infty}S_{n}^{\dagger}
    \left(\omega_{n}+i\frac{f}{2}\etagamma \right)S_{n}\, .
\end{equation}
The matrices $i\,\gamma^{2k-1,2k}$ are commuting matrices and have
eigenvalues $\pm 1$, each with multiplicity four. Since they
commute we can find a set of common eigenvectors. Choosing this set
as basis we can write the fermionic spectrum as
\begin{equation}\label{rotferH}
    H_{lc}^{F}= \frac{1}{ p^+}\sums{n=-\infty}{+\infty} \sum_{b=1}^{8}
    \left(\omega_n + \frac{f}{2} d_b \right)F_{n}^{(b)}\, ,
\end{equation}
where $F_{n}^{(b)}$ are the fermionic number operators defined by the relation
\begin{equation}
    F_{n}^{(b)}=\left(S_{n}^{b}\right)^{\dagger}S_{n}^{b}\,
\end{equation}
and where we have defined the coefficients $d_b$ as the following
combinations of the $\eta_k$ parameters
\begin{equation}
    \begin{array}{lll}
        d_1 = -\eta_{1}-\eta_{2}+\eta_{3}+\eta_{4} \, , \phantom{qquad} & d_5 = -\eta_{1}+\eta_{2}+\eta_{3}-\eta_{4} \, ,\\[1mm]
        d_2 = -\eta_{1}-\eta_{2}-\eta_{3}-\eta_{4} \, , & d_6 = \eta_{1}-\eta_{2}+\eta_{3}-\eta_{4} \, ,\\[1mm]
        d_3 = \eta_{1}+\eta_{2}+\eta_{3}+\eta_{4} \, ,  & d_7 = \eta_{1}-\eta_{2}-\eta_{3}+\eta_{4} \, ,\\[1mm]
        d_4 = \eta_{1}+\eta_{2}-\eta_{3}-\eta_{4} \, ,  & d_8 = -\eta_{1}+\eta_{2}-\eta_{3}+\eta_{4} \, .
    \end{array}
\end{equation}

At this point we can write the total light-cone Hamiltonian, $H_{lc}$, of  type
IIB string theory on the {\em rotated pp-wave \bg s}
\begin{equation}\label{eq:rotH}
\begin{split}
    H_{lc}=&H_{lc}^{B} +H_{lc}^{F}= \frac{1}{ p^+}\sum_{n=-\infty}^{+\infty}  \left\{\sum_{k=1}^2
    \left[\left(\omega_n + \eta_k f\right) M_{n}^{(k)}+\left(\omega_n - \eta_k f\right) \tilde{M}_{n}^{(k)}\right]\right. \\
    +&\left.\sum_{k=1}^2\left[\left(\omega_n + \eta_{(k+2)}  f\right) N_{n}^{(k)}+\left(\omega_n - \eta_{(k+2)}  f\right) \tilde{N}_{n}^{(k)}\right]
    + \sum_{b=1}^{8}\left(\omega_n + \frac{f}{2} d_b \right)F_{n}^{(b)}\right\}\, ,
\end{split}
\end{equation}
and the level matching condition is
\begin{equation}
    \sum_{n=-\infty}^{+\infty}\left[\sum_{k=1}^2\left(M_{n}^{(k)}+\tilde{M}_{n}^{(k)}
    +N_{n}^{(k)}+\tilde{N}_{n}^{(k)}\right)+ \sum_{b=1}^{8}F_{n}^{(b)}\right]=0 \, .
\end{equation}

Note that the spectrum does not depend on the $C_k$ parameters  since
they just represent a gauge choice, but only on the $\eta_k$ parameters.


\section{The decoupled sectors}
\label{sec:decsectors}

In this section we show that by taking a certain limit of the
spectra \eqref{eq:rotH}, we can reproduce the spectrum of anomalous dimensions of gauge theory operators in the dual sectors of $\mathcal{N}=4$ SYM theory found
in~\cite{Harmark:2007px}. The procedure follows that of \cite{Harmark:2006ta} where the spectrum in the $SU(2)$ sector is matched. Here we generalize this to all sectors that include scalar fields on the gauge theory side.

According to the AdS/CFT
correspondence, the string light-cone Hamiltonian $H_{\rm lc}$ should be dual to $D-J$ on the gauge theory side,
\begin{equation}
    \frac{H_{\rm lc}}{\mu}\, \longleftrightarrow \, D-J \, .
\end{equation}
where $D$ is the dilatation operator and $J$ is the total
charge defined by $J = n_1 S_1 + n_2 S_2 + n_3 J_1 + n_4 J_2 + n_5 J_3$ with the $n_i$ characterizing the decoupling limit giving a particular sector of $\CN=4$ SYM \cite{Harmark:2007px}. As explained in more detail below, the decoupling limit on the gauge theory consists of taking the limit $D-J \rightarrow 0$ and $\lambda\rightarrow 0$ keeping $(D-J)/\lambda$ fixed. On the string theory side, this decoupling limit corresponds to the limit $\mu \to \infty$, or
equivalently $f \to \infty$. We now apply this limit to the string spectra
\eqref{eq:rotH}. Remembering the definition of $\omega_n$, its
expansion for $f \to \infty$ takes the form
\begin{equation}
    \omega_n=\sqrt{f^2 + n^2}\simeq f+\frac{n^2}{2f} +\mathcal{O}(f^{-2})\, .
\end{equation}
In order for the spectra to be finite,
the divergent term contained in the expansion of $\omega_n$ should cancel.

In the bosonic part of the Hamiltonian \eqref{eq:rotH} we deal with terms of the kind
\begin{subequations}
    \begin{align}
        \left(\omega_n + \eta_k f\right)M_{n}^{(k)} &
        \simeq \left[f\left(1+\eta_k \right) + \frac{n^2}{2f}
        +\mathcal{O}(f^{-2})\right] M_{n}^{(k)}\, ,\\
        \left(\omega_n - \eta_k f\right)\tilde{M}_{n}^{(k)} &
        \simeq \left[f\left(1-\eta_k \right) + \frac{n^2}{2f}
        +\mathcal{O}(f^{-2})\right] \tilde{M}_{n}^{(k)}\, ,
    \end{align}
\end{subequations}
and the analogous ones for $N_{n}^{(k)}$ and $\tilde{N}_{n}^{(k)}$.
Instead in the fermionic part of the Hamiltonian \eqref{eq:rotH} we have
\begin{equation}
    \left(  \omega_n + \frac{f}{2} d_b\right)F_{n}^{(b)} \simeq \left[f\left(1+\frac{d_b}{2}\right)
    + \frac{n^2}{2f} +\mathcal{O}(f^{-2})\right]F_{n}^{(b)}\, .
\end{equation}
The only terms that survive the limit $f \to \infty$ are those for which the coefficient
of the linear part in $f$ vanishes.
All the other terms are divergent and thus decouple in the large $f$ limit.
The bosonic number operators
will survive only if the corresponding $\eta_k$ results to be $\pm 1$
and the fermionic number operators only if the corresponding
$d_b$ results to be $-2$.

In the following we want to show that by appropriately fixing the
values of the parameters $\eta_k$, the string theory spectra that
survive the limit $\mu \to \infty$ precisely reproduce the spectra
of the dual gauge theory sectors. As an important consequence of the
matching of the spectra, it follows that also the Hagedorn
temperature of the gauge theory matches the one of string theory in
these sectors. This can also be used to verify the conjectured
relation between the Hagedorn/deconfinement temperature of planar
$\CN=4$ SYM on $\R\times S^3$ and the Hagedorn temperature of string
theory on $\mbox{AdS}_5\times S^5$. Moreover, these results show that
the decoupling
limits~\cite{Harmark:2006di,Harmark:2006ta,Harmark:2006ie,Harmark:2007et,Harmark:2007px}
of thermal $SU(N)$ $\mathcal{N}=4$ SYM on $\R\times S^3$ provide a
very useful and powerful tool to  match gauge theory and string
theory.

On the gauge theory side the
idea~\cite{Harmark:2006di,Harmark:2006ta,Harmark:2006ie,Harmark:2007et,
Harmark:2007px, Harmark:2008gm} is to consider decoupling limits of
weakly coupled $\CN=4$ SYM on $\R \times S^3$ with gauge
group $SU(N)$.  The decoupling
limit is defined by
\begin{equation}
\label{limit2} \lambda \rightarrow 0 \spa J_i,\, N \ \mbox{fixed}
\spa H_{\rm g.t.} \equiv \frac{E-J}{\lambda} \ \mbox{fixed}
\end{equation}
where $\lambda=\gym^2 N/4\pi^2$ is the 't Hooft
coupling of $\mathcal{N}=4$ SYM theory, $E$ is the energy of a state
measured in units of the three sphere radius and $J\equiv n_1 S_1 +
n_2 S_2 + n_3 J_1 + n_4 J_2 + n_5 J_3$ is the total charge with
$n_i$, $i=1,\ldots,5$ being fixed numbers. $S_1$ and $S_2$ denote
the two charges of the $SO(4)$ group of $S^3$ and $J_1$, $J_2$ and
$J_3$ are the three R-charges. Here we only consider the gauge
theory in the planar limit $N=\infty$. In terms of operators we have
that the Hamiltonian is given by $H_{\rm g.t.} = (D-J)/\lambda$. $D$
is the dilatation operator of $\mathcal{N}=4$ SYM which, at weak 't
Hooft coupling, can be expanded as
\begin{equation}
D = D_0 + \lambda D_2 + \lambda^{\frac{3}{2}}D_3 + \lambda^2D_4 + \ldots
\end{equation}
where $D_0$ is the bare scaling dimension, $D_2$ is the one-loop
part of the dilatation operator and so on. One can see that in the
limit \eqref{limit2}, the operators with $D_0>J$ decouple and only
the ones with $D_0=J$ survive the limit. One thus gets the effective
Hamiltonian $H_{\rm g.t.}=D_2$, namely only the one-loop part of the
dilatation operator survive the limit
\eqref{limit2}~\cite{Harmark:2006di,Harmark:2006ta,Harmark:2006ie,Harmark:2007et,Harmark:2007px,
Harmark:2008gm}.

Among the possible decoupling limits of $\mathcal{N}=4$ SYM theory
found in~\cite{Harmark:2007px}, here we are interested only in the
decoupled sectors that contain scalars. The presence of the scalars
is in fact crucial in order to analyze the regime of the gauge
theory which is related to the dual string theory. These sectors are
the $SU(2)$, $SU(1|1)$, $SU(1|2)$, $SU(2|3)$, bosonic $SU(1,1)$,
$SU(1,1|1)$, $SU(1,1|2)$, $SU(1,2|2)$ and $SU(1,2|3)$ sectors.
\begin{table}[ht]
\begin{center}
\begin{tabular}{|l||c|}
  \hline
   \textbf{Sector} &  $(n_1,n_2,n_3,n_4,n_5)$ \\
    \hline  \hline
     $SU(2)$ &  (0,0,1,1,0) \\
    \hline
     $SU(1,1)_{b}$  &(1,0,1,0,0)\\
    \hline
     $SU(1|1)$  & $\left(\frac{2}{3},0,1,\frac{2}{3},\frac{2}{3}\right)$ \\
    \hline
    $SU(1|2)$  & $\left(\frac{1}{2},0,1,1,\frac{1}{2}\right)$ \\
    \hline
     $SU(2|3)$  & (0,0,1,1,1) \\
    \hline
     $SU(1,1|1)$  & $\left(1,0,1,\frac{1}{2},\frac{1}{2}\right)$ \\
    \hline
     $SU(1,1|2)$  & (1,0,1,1,0) \\
    \hline
     $SU(1,2|2)$  & (1,1,1,0,0) \\
    \hline
     $SU(1,2|3)$  & (1,1,1,1,1) \\
    \hline
\end{tabular}
\caption{The table shows the nine decoupled sectors that contain at
least one scalar: in the left column are listed the sectors that
survive the decoupling limit for the corresponding choice of
$n=(n_1,n_2,n_3,n_4,n_5)$ reported in the right column. $SU(1,1)_b$
is the bosonic $SU(1,1)$ sector.} \label{tab:sectors}
\end{center}
\end{table}
For more details see Ref.~\cite{Harmark:2007px}.

The spectra for these nine different sectors all take the
form~\cite{Harmark:2007px}
\begin{equation}
\label{eq:ABCspectrum2} H_{\rm g.t.}  =
 \frac{2\pi^2}{J^2} \sum_{n\in \mathbb{Z}} n^2
\left( \sum_{i=1}^a M_n^{(i)} +\sum_{j=1}^b N_n^{(j)} +
\sum_{\alpha=1}^c F_n^{(\alpha)} \right)
\end{equation}
The cyclicity (zero momentum) constraint is
\begin{align}
\label{eq:ABCconstraint} P \equiv \sum_{n\in \mathbb{Z}} n \left(
\sum_{i=1}^a M_n^{(i)} +\sum_{j=1}^b N_n^{(j)} + \sum_{\alpha=1}^c
F_n^{(\alpha)}  \right) = 0.
\end{align}
Note that $F_n^{(\alpha)} \in \{0,1\}$ while $M_n^{(i)}, N_n^{(j)}
\in \{0,1,2,...\}$. The numbers $a,b$ and $c$ are given in
Tab.~\ref{tab:abc}.
\begin{table}
\begin{center}
\begin{tabular}{c|ccccccccc}
$SU(\cdot)$&$(2)$ & $(1,1)_b$ & $(1|1)$ & $(1|2)$ & $(2|3)$ & $(1,1|1)$ & $(1,1|2)$ & $(1,2|2)$ & $(1,2|3)$ \\
\hline
    $a$ & 1 & 0 & 0 & 1 & 2 & 0 & 1 & 0 & 2 \\
    $b$ & 0 & 1 & 0 & 0 & 0 & 1 & 1 & 2 & 2 \\
    $c$ & 0 & 0 & 1 & 1 & 2 & 1 & 2 & 2 & 4
\end{tabular}
\caption{The table shows how many number operators we have of each
type ($a$ for scalars $M_n$, $b$ for derivatives $N_n$, and $c$ for
fermions $F_n$) in each of the nine theories that contain at least
one scalar. $SU(1,1)_b$ is the bosonic $SU(1,1)$ sector.
\label{tab:abc}}
\end{center}
\end{table}

We want to show that there is a direct relation between the critical
values of the numbers $(n_1,...,n_5)$ that characterize the various
sectors on the gauge theory side and the parameters
$\eta_1,...,\eta_4,$ that give the corresponding decoupled sectors
on the string theory side.

From table \ref{tab:sectors}, we see that all the nine sectors
containing scalars have $n_3 = 1$. It is not hard to see that a
suitable choice of $\eta_k$ parameters to match the string theory spectrum
with the spectrum of the gauge theory side is the following
\begin{equation}\label{eq:etaasn}
        \eta_1 =n_4 \, ,\phantom{qquad}  \eta_2 =n_5 \, , \phantom{qquad}
        \eta_3 =-n_1 \, ,\phantom{qquad}  \eta_4 =n_2 \, .
\end{equation}
Using the previous relations in the spectrum \eqref{eq:rotH}
and taking the limit $f\to \infty$ we see that the
string theory spectrum precisely matches the spectrum of the
nine decoupled sectors of the gauge theory side.

As an example we can consider the $SU(1,1|1)$ sector: in this case
$n=\left(1,0,1,\frac{1}{2},\frac{1}{2}\right)$ (see Table
\ref{tab:sectors}) so using the relations \eqref{eq:etaasn} we have
that $\eta=\left(\frac{1}{2},\frac{1}{2},-1,0\right)$. Since the
only $\eta_k$ equal to -1 is $\eta_3$ and the only $d_b$ equal to -2
is $d_1$ we have that only one bosonic and one fermionic number
operator survive the limit $f \to \infty$. The string theory
spectrum thus becomes
\begin{equation}\label{strsect2}
        \frac{H_{lc}}{\mu}\sim \frac{1}{2 \mu p^+ f} \sum_{n\in \mathbb{Z}} n^2 \left( N_n^{(1)}
        + F_n^{(1)} \right)\, ,
\end{equation}
which, using the dictionary between gauge theory and string theory,
can be written as
\begin{equation}\label{strsect}
        \frac{H_{lc}}{\mu}= \lambda D_2=\frac{2\pi^2\lambda}{J^2} \sum_{n\in \mathbb{Z}} n^2 \left( N_n^{(1)}
        + F_n^{(1)} \right)\, ,
\end{equation}
where we used that $f=J/(2\pi\sqrt{\lambda})$. It is easy to check
that \eqref{strsect} is in accordance with the corresponding result
in the gauge theory side which can be deduced from
\eqref{eq:ABCspectrum2}.

We can repeat an analogous check for all the other decoupled sectors
and we can show that the field content of the surviving spectrum is
exactly the same as the one obtained on the gauge theory side.

Using again Table \ref{tab:abc}, we can thus write the reduced spectrum
for all the nine sectors on the string theory side at once. It is given by
\begin{equation}
\frac{H_{lc}}{\mu}=\frac{1}{2 \mu p^+ f} \sum_{n\in \mathbb{Z}}
n^2 \left( \sum_{i=1}^a M_n^{(i)} +\sum_{j=1}^b N_n^{(j)} +
\sum_{\alpha=1}^c F_n^{(\alpha)} \right)
\end{equation}
which indeed coincides with Eq. \eqref{eq:ABCspectrum2} once we use the dictionary between gauge theory and string theory.

\section{New Penrose limit of $\ads_4 \times \C P^3$}
\label{sec:ads4}

In the above we have found a new Penrose limit of $\ads_5 \times S^5$ with two explicit space-like isometries in addition to the existing Penrose limits with zero and one space-like isometries \cite{Blau:2002dy,Berenstein:2002jq,Bertolini:2002nr}. A natural question is whether one can similarly find new Penrose limits of the $\ads_4\times \C P^3$ background of type IIA supergravity. The known Penrose limits for this background are with either zero explicit space-like isometries \cite{Nishioka:2008gz,Gaiotto:2008cg} or with two space-like isometries \cite{Grignani:2008is,Astolfi:2009qh}. In particular the one with two space-like isometries of \cite{Grignani:2008is,Astolfi:2009qh} is connected to studying the $SU(2) \times SU(2)$ sector of string theory on $\ads_4\times \C P^3$.

We find in this section a new Penrose limit of the $\ads_4\times \C P^3$ background of type IIA supergravity with one explicit space-like isometry, $i.e.$ with one flat direction. We find furthermore the spectrum of type IIA string theory on this background by finding the spectrum for a general rotated pp-wave background that for certain choices of parameters corresponds to both the new pp-wave background with one explicit space-like isometry, as well as the two known backgrounds with zero and two explicit space-like isometries.

\subsection{The ``one flat direction'' Penrose limit}

In this section we present a new Penrose limit of $\ads_4 \times \C P^3$, here called the ``one flat direction'' Penrose limit.
The \adscp\ metric is given by
\begin{equation}
	ds^2=R^2\left(\frac{1}{4}ds^2_{AdS_4}+ ds^2_{\CP^3}\right) \, ,
\end{equation}
where
\begin{equation}\label{metricAdS4}
	ds^2_{AdS_4} =
	-\cosh^2\rho \, dt^2 +d\rho^2 +\sinh^2 \rho \, d\Omega_2^2 \, ,
\end{equation}
and
\begin{equation}\label{metricCP3}
	\begin{split}
		ds^2_{\CP^3} & = d\theta^2+4\cos^2 \theta \sin^2 \theta \left(d\delta+\frac{\cos\theta_1}{4}d\vp_1-
		\frac{\cos\theta_2}{4}d\vp_2\right)^2 \\
		&+\frac{1}{4}\cos^2 \theta\left(d\theta_1^2+\sin^2\theta_1
		d\vp_1^2\right)+\frac{1}{4}\sin^2 \theta (d\theta_2^2+\sin^2\theta_2
		d\vp_2^2)\, .
	\end{split}
\end{equation}
We introduce the new variables $\chi$, $\xi$ and $\psi$ by
\begin{equation}
	2\delta = \chi + \frac{\vp_2}{2}\, ,\qquad \vp_2=\xi+b\chi\,, \qquad 2\theta = \psi+ \frac{\pi}{2}\, ,
\end{equation}
where $b$ is a parameter.
The coordinate transformation that defines the Penrose limit is
\begin{equation}
		\begin{split}
		&x^+ = \frac{t+\chi}{2} \,, \qquad x^- = R^2\frac{t-\chi}{8} \,,  \qquad \rho= \frac{2r}{R} \, ,\qquad \psi=\frac{2 u_4}{R} \, ,\\
		&\vp_1  = \frac{2\sqrt{2}\,x_1}{R} \, ,\qquad \theta_1=\frac{2\sqrt{2}\,y_1}{R}+\frac{\pi}{2} \, ,\qquad
		\theta_2=\frac{2\sqrt{2}\,z}{R}\, .
	\end{split}
\end{equation}
Taking the limit $R \to \infty$ while keeping $x^\pm$, $r$, $u_4$, $x_1$, $y_1$, $z$ finite,
the metric becomes
\begin{equation}\label{metric1fd}
	\begin{split}
		ds^2 = &-4dx^+  dx^- + \sums{i=1}{4}\left(du_i^2-u_i^2 {dx^+}^2\right) + \sum_{a=1}^{2}\left(dx_a^2+dy_a^2 \right) \\
		&+b(1+b)\left(x_2^2+y_2^2\right){dx^+}^2- 2 y_1 dx_1 dx^+
		+(1+2b)\left[x_2 dy_2 - y_2 dx_2\right]dx^+ \, ,
	\end{split}
\end{equation}
where $x_2+iy_2=z\,e^{i\xi}$. The metric \eqref{metric1fd} describes exactly a
 a pp-wave \bg\ with a flat direction, namely $x_1$.

\subsection{Rotated \bg s and the string spectrum}

Let us start from the pp-wave metric found in \cite{Nishioka:2008gz}
\begin{equation}
	ds^2=-4d\tilde{x}^+d\tilde{x}^- -\left(\sum_{i=1}^4
\tilde{x}_i^2+\frac{1}{4}\sum_{i=5}^8\tilde{x}_i^2\right){d\tilde{x}^+}^2+\sum_{i=1}^8 d\tilde{x}_i^2,
\end{equation}
We consider the following coordinate transformation
\begin{equation}\label{transfrot2}
\begin{split}
\tilde x^+ &=x^+  \, \\[2mm]
\tilde x^- &=x^- + \sum_{a=1}^{2} C_a x_a y_a \, , \\[2mm]
\tilde x_i &=u_i \, , \quad i=1,\dots,4\, ,\\[2mm]
\left( \begin{array}{c}
 \tilde x_{3+2a} \\[2mm]
 \tilde x_{4+2a}
\end{array} \right) &=
\left( \begin{array}{cc}
 \cos(\eta_a x^+ ) & -\sin(\eta_a x^+ ) \\[2mm]
 \sin(\eta_a x^+ ) & \cos(\eta_a x^+ )
\end{array} \right)
\left( \begin{array}{c}
 x_a \\[2mm]
 y_a
\end{array} \right)\, , \quad a=1,2\, ,
\end{split}
\end{equation}
where  $C_{1}, C_{2}$ and $\eta_{1}, \eta_{2}$ are parameters.
Under this tranformation the metric becomes
\begin{align}\label{rotmet}
	ds^2 =& -4dx^+   dx^- + \sums{i=1}{4}\left(du_i^2-u_i^2 {dx^+ }^2\right)
	+ \sum_{a=1}^{2}\Bigg[dx_a^2+dy_a^2+\left(\eta_a^2-\frac{1}{4}\right)\left(x_a^2+y_a^2\right){dx^+ }^2 \nn \\
	& +2\left(\eta_a-2C_a\right)x_a dy_a dx^+  - 2\left(\eta_a+2C_a\right)y_a dx_a dx^+ \Bigg]\, .
\end{align}
It is easy to see that
if one chooses the $C_a$ and $\eta_a$ parameters so the terms $dx^+ {}^2$ and $dx_a dx^+ $ in the metric
\eqref{rotmet} vanish, i.e.
\begin{equation}
		\eta_a=-\frac{1}{2}\, ,\qquad \quad C_a=\frac{1}{4} \, ,
\end{equation}
then one gets the \bg\ with two flat directions found in \cite{Grignani:2008is}
\begin{equation}
	ds^2 = -4dx^+   dx^- + \sums{i=1}{4}\left(du_i^2-u_i^2 {dx^+ }^2\right) + \sum_{a=1}^{2}\left[dx_a^2+dy_a^2 - 2 y_a dx_a dx^+ \right]
\end{equation}

Eq.~\eqref{rotmet} also contains the pp-wave \bg\
with one flat direction \eqref{metric1fd} that we just obtained
through a Penrose limit of the \adscp\
geometry for the following choice of parameters
\begin{equation}
	\begin{array}{lcl}
		\eta_1=- \displaystyle \frac{1}{2} \, , & \phantom{aaa}& \eta_2=b+  \displaystyle \frac{1}{2} \, , \\[2mm]
		 C_1= \displaystyle \frac{1}{4} \, ,& &  C_2=0 \, .
	\end{array}
\end{equation}

\subsubsection*{Spectrum}
Now we derive the string spectrum on the rotated pp-wave \bg\ \eqref{rotmet}.

In the light-cone gauge $x^+ = c \tau$ the bosonic Lagrangian density is
\begin{equation}\label{penboslagr}
\begin{split}
    &\mathscr{L}_{\rm lc}^{B}= - \frac{1}{4\pi  c}\bigg\{\sum_{i=1}^4\left[\dot{u}_i^2
    -u_i'^2-c^2u_i^2\right]+\sum_{a=1}^2\Big[\dot{x}_a^2+\dot{y}_a^2
    -x_a'^2-y_a'^2 \\
    &+c^2\left(\eta_a^2-\frac{1}{4}\right)\left(x_a^2+y_a^2\right)+2c\left(\eta_a-2 C_a\right)x_a \dot{y}_a
    -2c\left(\eta_a+2C_a\right)y_a \dot{x}_a\Big]\bigg\}\, .
\end{split}
\end{equation}
where $c$ is fixed by requiring that the conjugate momentum to $x^-$ is constant.
The bosonic light-cone Hamiltonian is then given by
\begin{equation} \label{penbosham}
\begin{split}
c H^B_{\rm lc}=&  \frac{1}{4\pi } \int_0^{2\pi} d\sigma
\bigg\{\sum_{i=1}^4\left[\dot{u}_i^2
 +u_i'^2+c^2u_i^2\right] \\
 &+\sum_{a=1}^2\left[\dot{x}_a^2+\dot{y}_a^2
 +x_a'^2+y_a'^2+c^2\left(\frac{1}{4}-\eta_a\right)\left(x_a^2+y_a^2\right)\right] \bigg\}\, .
\end{split}
 \end{equation}
The mode expansion for the bosonic fields can be written as
\begin{equation}
u_i (\tau,\sigma ) = \frac{i}{\sqrt{2}} \sum_{n\in \Z}
\frac{1}{\sqrt{\Omega_n}} \Big[ \hat{a}^i_n e^{-i ( \Omega_n \tau -
n \sigma ) } - (\hat{a}^i_n)^\dagger e^{i ( \Omega_n \tau - n \sigma
) } \Big] \, ,
\end{equation}
\begin{equation}\label{zmode}
z_a(\tau,\sigma) =  \, e^{-i c \eta_a
\tau} \sum_{n \in \Z} \frac{1}{\sqrt{\omega_n}} \Big[ a_n^a
e^{-i ( \omega_n \tau - n \sigma ) } -  (\tilde{a}^a)^\dagger_n e^{i
( \omega_n \tau - n \sigma ) } \Big]\, ,
\end{equation}
where $\Omega_n=\sqrt{c^2+n^2}$, $\omega_n=\sqrt{\frac{c^2}{4}+n^2}$
and we defined $z_a(\tau,\sigma)=x_a(\tau,\sigma)+iy_a(\tau,\sigma)$.
The canonical
commutation relations $[x_a(\tau,\sigma),p_{x_b}(\tau,\sigma')] =
i\delta_{ab} \delta (\sigma-\sigma')$,
$[y_a(\tau,\sigma),p_{y_b}(\tau,\sigma')] = i\delta_{ab}\delta
(\sigma-\sigma')$ and $[u_i(\tau,\sigma),p_j(\tau,\sigma')] =
i\delta_{ij} \delta (\sigma-\sigma')$ follows from
\begin{equation}
\label{comrel} [a_m^a,(a_n^b)^\dagger] = \delta_{mn} \delta_{ab}\spa
[\tilde{a}_m^a,(\tilde{a}_n^b)^\dagger] = \delta_{mn}
\delta_{ab}\spa [\hat{a}^i_m,(\hat{a}^j_n)^\dagger] = \delta_{mn}
\delta_{ij} \, .
\end{equation}
Employing \eqref{comrel} we obtain the bosonic spectrum
\begin{equation}\label{penspectrum}
c H^B_{\rm lc} = \sum_{i=1}^4 \sum_{n\in \Z} \sqrt{n^2+c^2}\,
\hat{N}^i_n+\sum_{a=1}^2\sum_{n\in \Z} \left\{
\left(\sqrt{\frac{c^2}{4}+n^2} + \eta_a c \right) M_n^a + \left(\sqrt{\frac{c^2}{4}+n^2}-
\eta_a c\right) N_n^a \right\} \, ,
\end{equation}
with the number operators $\hat{N}^i_n = (\hat{a}^i_n)^\dagger
\hat{a}^i_n$, $M_n^a = (a^a)^\dagger_n a^a_n$ and $N_n^a =
(\tilde{a}^a)^\dagger_n \tilde{a}_n^a$.

Now we compute the fermionic part of the spectrum. We start from the type IIA
superstring Lagrangian density on the \bg\ \eqref{rotmet}
\begin{equation}\label{penferlagr}
    \mathscr{L}^{F}= \frac{i \,c }{2} \, \bar{\theta} \Gamma_+ \left[\partial_\tau -\Gamma_{11} \partial_\sigma
    +\frac{c}{4}\left(-2\eta_1\Gamma_{56}-2\eta_2\Gamma_{78}+\Gamma_{11}\Gamma_4-3\Gamma_{123}\right)\right]\theta \, ,
\end{equation}
where $\theta$ is a 32 component real spinor and we used the zehnbeins
\begin{equation}
\begin{split}
	&e^+_{\phantom{+}+}=\frac{1}{2} \, , \qquad e^-_{\phantom{+}+}=\frac{1}{2}\left[\left(\sums{i=1}{4}u_i^2\right) -
	\sum_{a=1}^{2} \left(\eta_a^2-\frac{1}{4}\right) \left(x_a^2+y_a^2\right)\right] \, ,\\
	&e^-_{\phantom{+}-}=2\, , \qquad e^-_{\phantom{+}x_a}=\left(\eta_a+2C_a\right)y_a \, ,
	\qquad e^-_{\phantom{+}y_a}=-\left(\eta_a-2C_a\right)x_a \, , \\
	&e^i_{\phantom{+}u_i}=1 \, , \qquad e^5_{\phantom{+}x_1}=1\, , \qquad e^6_{\phantom{+}y_1}=1\, ,
	 \qquad e^7_{\phantom{+}x_2}=1\, , \qquad e^8_{\phantom{+}y_2}=1\, ,
\end{split}
\end{equation}
where $i=1,2,3,4$, and the relevant components of the spin connection
\begin{equation}
	\omega_+^{\phantom{+}56}=-\eta_1\, , \qquad \omega_+^{\phantom{+}78}=-\eta_2\, .
\end{equation}

Let us decompose $\theta=\theta_+ +\theta_-$ by writing
\begin{equation}
	\Gamma_{5678}\theta_\pm=\pm \theta_\pm  \, ,
\end{equation}
In terms of $\theta_\pm$ the light-cone gauge conditions are \cite{Astolfi:2009qh}
\begin{equation}
	\Gamma_- \theta_- =0\, , \qquad \Gamma_{4956}\theta_+=\theta_+\, .
\end{equation}
Using the spinor conventions of Appendix \ref{AppendixA} we can write the Lagrangian as
\begin{equation}
	\mathscr{L}^{F} = \mathscr{L}_+ +\mathscr{L}_- \, ,
\end{equation}
with $\mathscr{L}_+$ and $\mathscr{L}_-$ given by
\begin{equation}
	\mathscr{L}_+=i \psi^* \dot{\psi}  - \frac{i}{2}\left(\psi \psi' + \psi^* {\psi^*}'\right) +\frac{i\, c}{2}\Delta_1  \psi \gamma_{56} \psi^*
 + \frac{c}{2} \psi \psi^* \, ,
\end{equation}
\begin{equation}
	\mathscr{L}_-=i \chi^* \dot{\chi}  - \frac{i}{2}\left(\chi \chi' + \chi^* {\chi^*}'\right) -\frac{i\, c}{2}\Delta_2  \chi \gamma_{56} \chi^*
 - c \chi \chi^* \, ,
\end{equation}
where $\Delta_1=\eta_2-\eta_1$ and $\Delta_2=\eta_1+\eta_2$.

The mode expansions for the 8 component spinors $\psi$ and  $\chi$ are
\begin{equation}
	\psi_{\alpha} =   \left( e^{- \frac{c}{2} \Delta_1 \gamma_{56}\tau} \right)_{\alpha \beta}
	\sum_{n\in Z} \left[ f^+_n d_{n,\alpha}e^{-i ( \omega_n \tau - n \sigma ) } - f^-_n d^\dagger_{n,\alpha}
	e^{i ( \omega_n \tau - n \sigma ) } \right]\, ,
\end{equation}
\begin{equation}
	\chi_{\alpha} =  \left( e^{ \frac{c}{2} \Delta_2 \gamma_{56}\tau} \right)_{\alpha \beta} \sum_{n\in Z} \left[ - g^-_n b_{n,\beta}
	e^{-i ( \Omega_n \tau - n \sigma ) } + g^+_n b^\dagger_{n,\beta}
	e^{i ( \Omega_n \tau - n \sigma ) } \right] \, ,
\end{equation}
with the constants $f^\pm_n$ and $g^\pm_n$ defined by
\begin{equation}
f^\pm_n = \frac{\sqrt{\omega_n+n} \pm
\sqrt{\omega_n-n}}{2\sqrt{\omega_n}} \spa g^\pm_n =
\frac{\sqrt{\Omega_n+n} \pm \sqrt{\Omega_n-n}}{2\sqrt{\Omega_n}}
\end{equation}
The fermionic Hamiltonian density is therefore
\begin{equation}\label{CH2F}
 c \mathcal{H}^{F}_{\rm lc} =
 \frac{i}{2} \left( \psi \psi' -\rho \rho' \right) + \frac{c}{2} \Delta_1 \psi \gamma_{56}\rho - \frac{i\, c}{2} \psi \rho 
  + \frac{i}{2}\left(\chi \chi' - \lambda \lambda'\right) - \frac{i\,c}{2} \Delta_2 \chi \gamma_{56}\lambda + i c \chi \lambda \, ,
\end{equation}
where the fermionic momenta are
\begin{equation}
\rho = - i \psi^* \spa \lambda = - i \chi^* \, .
\end{equation}
The fermionic spectrum can then be computed and reads
\begin{equation}\label{fermppwave}
\begin{split}
	c H^{F}_{\rm lc} &= \sum_{n\in \Z} \Bigg[ \sum_{b=1,2}\left(\omega_n +\frac{c}{2} \Delta_1 \right)F_n^{(b)} +
	\sum_{b=3,4}\left(\omega_n -\frac{c}{2} \Delta_1 \right)F_n^{(b)} \\
	&+	\sum_{b=5,6} \left( \Omega_n - \frac{c}{2}\Delta_2 \right) F_n^{(b)} +
	\sum_{b=7,8} \left( \Omega_n +   \frac{c}{2}\Delta_2 \right) F_n^{(b)} \Bigg]
\end{split}
\end{equation}
with the number operators $F^{(b)}_n= d_{n,\alpha}^\dagger d_{n,\alpha}$ for $b=1,\ldots ,4$, and
$F_n^{(b)} = b^\dagger_{n,\alpha} b_{n,\alpha}$ for $b=5,\ldots,8$. The level-matching condition, including also the bosonic part, is
\begin{equation}
\label{levelmbf} \sum_{n\in \Z}n \left[\sum_{i=1}^4
\hat{N}^i_n+\sum_{a=1}^2 \left(M_n^a + N_n^a\right) +\sum_{b=1}^8
F^{(b)}_n\right]
 = 0
\end{equation}

\section*{Acknowledgments}

GG and AM thank the Galielo Galilei Institute for Theoretical Physics for hospitality and the INFN for partial support during the completion of this work. The work of GG is supported in part by the MIUR-PRIN contract 2007-5ATT78.
\begin{appendix}

\section{Gamma matrices and spinors}\label{AppendixA}

We briefly review our conventions for the representations of Dirac
matrices in ten dimensions and for Majorana-Weyl spinors.
As usual, we shall use the mostly plus metric.

\subsection*{Gamma matrices}

Let $I_n$ denote the $n \times n$ unit matrix,  $\sigma_1,\, \sigma_2,\, \sigma_3$
the $2\times 2$ Pauli matrices
\begin{equation}\label{Paulimatr}
    \sigma_1 = \matrto{0}{1}{1}{0}\lspa \sigma_2 = \matrto{0}{-i}{i}{0} \lspa
    \sigma_3 = \matrto{1}{0}{0}{-1}\, ,
\end{equation}
and $\epsilon$ the antisymmetric tensor of rank two
\begin{equation}
\epsilon = i \sigma_2 =\matrto{0}{1}{-1}{0} \, .
\end{equation}

We can define the real $8 \times 8$ matrices $\gamma_1,...,\gamma_8$ as
\begin{equation}\label{gammadirac}
    \begin{array}{ll}
    \gamma_1 = \epsilon \times \epsilon \times \epsilon\, , \phantom{qquad} & \gamma_5 = \sigma_3 \times \epsilon \times I_2\, , \\[1mm]
    \gamma_2 = I_2 \times \sigma_1 \times \epsilon \, , & \gamma_6 =\epsilon \times I_2 \times \sigma_1 \, , \\[1mm]
    \gamma_3 = I_2 \times \sigma_3 \times \epsilon \, ,& \gamma_7 = \epsilon \times I_2 \times \sigma_3 \, , \\[1mm]
  \gamma_4 = \sigma_1 \times \epsilon \times I_2 \, , & \gamma_8 = I_2\times I_2 \times I_2\, .
    \end{array}
\end{equation}
This should be read as
\begin{equation}\label{notgammamatr}
\gamma_7 = \epsilon \times I_2 \times \sigma_3 = \matrto{0}{I_2
\times \sigma_3}{-I_2 \times \sigma_3}{0} \spa I_2 \times \sigma_3 =
\matrto{\sigma_3}{0}{0}{\sigma_3}\, ,
\end{equation}
and so on. It is easy to verify that the matrices $\gamma_1,...,\gamma_8$ obey
the following relations
\begin{equation}\label{smallgamma9}
    \begin{split}
    &\gamma_i \gamma_j^T + \gamma_j \gamma_i^T = \gamma_i^T \gamma_j +
    \gamma_j^T \gamma_i = 2 \delta_{ij} I_8 \lspa i,j=1,...,8 \\[1mm]
    & \gamma_1 \gamma_2^T \gamma_3 \gamma_4^T \gamma_5
    \gamma_6^T \gamma_7 \gamma_8^T = I_8 \lspa \gamma_1^T \gamma_2
    \gamma_3^T \gamma_4 \gamma_5^T \gamma_6 \gamma_7^T \gamma_8 = - I_8\, .
    \end{split}
\end{equation}
Now we introduce the $16 \times 16$ matrices $\hat{\gamma}_1,...,\hat{\gamma}_9$ defined as
\begin{equation}\label{gam9}
    \begin{split}
    &\hat{\gamma}_i = \matrto{0}{\gamma_i}{\gamma_i^T}{0}\, , \qquad
    i,j=1,...,8 \\[1mm]
    &\hat{\gamma}_{9} = \sigma_3 \times I_8 =
    \matrto{I_8}{0}{0}{-I_8}\, .
    \end{split}
\end{equation}
The matrices $\hat{\gamma}_1,...,\hat{\gamma}_9$
are symmetric and real, and they obey
\begin{equation}
    \begin{split}
    \{ \hat{\gamma}_i, \hat{\gamma}_j \} &= 2 \delta_{ij} I_{16} \, , \qquad
    i,j=1,...,9 \\[1mm]
    &\hat{\gamma}_9 = \hat{\gamma}_1 \hat{\gamma}_2 \cdots \hat{\gamma}_8 \, .
    \end{split}
\end{equation}

At this point we are ready to define the Dirac matrices in ten dimensions,
which are the following $32 \times 32$ matrices:
\begin{equation}
    \begin{split}
    \Gamma_0 &= - \epsilon \times I_{16} = \matrto{0}{-I_{16}}{I_{16}}{0} \, , \\
    \Gamma_i &= \sigma_1 \times \hat{\gamma}  =
    \matrto{0}{\hat{\gamma}_i}{\hat{\gamma}_i}{0} \, , \quad i=1,...,9 \\
    \Gamma_{11} &= \sigma_3 \times I_{16} =
    \matrto{I_{16}}{0}{0}{-I_{16}}\, .
    \end{split}
\end{equation}
We see that these matrices are real and satisfy the relations
\begin{equation}\label{gam11}
    \begin{split}
    \{ \Gamma_a,\Gamma_b \} = 2 \eta_{ab} I_{32} \, ,& \quad
    a,b=0,1,...,9,11 \\[1mm]
    \Gamma_{11} = \Gamma^0 & \Gamma^1 \cdots \Gamma^9 \, .
    \end{split}
\end{equation}
It is convenient to introduce the light-cone Dirac matrices $\Gamma_\pm$,
given by
\begin{equation}
    \begin{split}
    \Gamma_\pm = &\Gamma_0 \pm \Gamma_9 \, , \\
    \Gamma^\pm = - \frac{1}{2} \Gamma_\mp &= \frac{1}{2} ( \Gamma^0 \pm
    \Gamma^9 )\, .
    \end{split}
\end{equation}
The raising and lowering of these indices are done according to a
flat space metric with $\eta_{+-} = -2$. \\

We then define
\begin{equation}
\Gamma_{a_1 a_2 \cdots a_n} = \Gamma_{[a_1} \Gamma_{a_2} \cdots
\Gamma_{a_n]} \, ,
\end{equation}
and analogously the $16 \times 16$ matrices
\begin{equation}
\hat{\gamma}_{i_1 \cdots i_n } = \hat{\gamma}_{[i_1}
\hat{\gamma}_{i_2} \cdots \hat{\gamma}_{i_n]} \, ,
\end{equation}
with $i_l = 1,...,8$. Since $\hat{\gamma}_i$ is symmetric we have
that
\begin{equation}
\hat{\gamma}_{ijkl}^T = \hat{\gamma}_{ijkl} \, ,
\end{equation}
i.e. that $\hat{\gamma}_{ijkl}$ is also symmetric.

Furthermore we define the $8\times 8$ matrices
\begin{equation}\label{gammai1ik}
\gamma_{i_1 \cdots i_{2k} } = \gamma_{[i_1} \gamma^T_{i_2}\cdots
\gamma^T_{i_{2k}]} \spa \gamma_{i_1 i_2 \cdots i_{2k+1} } =
\gamma^T_{[i_1} \gamma_{i_2} \cdots \gamma^T_{i_{2k+1}]} \, .
\end{equation}
with $i_l = 1,...,8$. In particular we call $\Pi$ the matrix
\begin{equation}
\Pi \equiv \gamma_{1234} = \gamma_1 \gamma_2^T \gamma_3 \gamma_4^T \, ,
\end{equation}
which has the following proprieties
\begin{equation}
\label{piids1} \Pi^2 = I_8 \spa \Pi^T = \Pi \spa \Pi = \gamma_{5678} \, .
\end{equation}
The last equation follows from \eqref{smallgamma9}. Finally it is possible to show that
$\Pi$ satisfies the relations
\begin{equation}\label{piids2}
\Pi \gamma_{ij} = \gamma_{ij} \Pi = - \epsilon_{ijkl}
\gamma^{kl} \spa \Pi \gamma_{i'j'} = \gamma_{i'j'} \Pi = -
\epsilon_{i'j'k'l'} \gamma^{k'l'} \, ,
\end{equation}
with $i,j=1,2,3,4$ and $i',j'=5,6,7,8$.

\subsection*{Spinors for type IIB}

The spinors $\theta^A$ are 32-component Majorana-Weyl spinors. The
Majorana condition imposes that the 32 components of $\theta^A$ are
real. The Weyl condition is
\begin{equation}
\label{weylcond}
\Gamma_{11} \theta^A = \theta^A \, ,
\end{equation}
for both $A=1,2$. Note here that we choose the two spinors to have
the same chirality since we are considering type IIB string theory.
Using \eqref{gam11} we see that the Weyl condition means that only
the first 16 components of $\theta^A$ are non-zero, whereas the last
16 components are zero. We write therefore
\begin{equation}
\label{defpsi}
\theta^A = \vecto{\psi^A}{0} \, ,
\end{equation}
where $\psi^A$, $A=1,2$, are two real 16 component spinors.

The light-cone gauge $\Gamma_- \theta^A = 0$
results to be equivalent to
\begin{equation}
\hat{\gamma}_9\psi^A = \psi^A \, ,
\end{equation}
which resembles a Weyl condition for the transverse directions.
Indeed, using \eqref{gam9}, we see that the last 8 components of
$\psi^A$ are zero. Thus, we write
\begin{equation}
\label{defS}
\psi^A = \vecto{S^A}{0} \, ,
\end{equation}
where $S^A$, $A=1,2$, are two real 8 component spinors.

\subsection*{Spinors for type IIA}

For the type IIA GS string we have two Majorana-Weyl spinors $\theta^{1,2}$ with opposite chirality,
$i.e.$ $\Gamma_{11} \theta^1 = \theta^1$ and $\Gamma_{11} \theta^2 = - \theta^2$. We collect these
into a 32 component real spinor $\theta = \theta^1 + \theta^2$.

We can then decompose $\theta$ in terms of eigenstates of $\Gamma_{5678}$ namely $\theta=\theta_+
+\theta_-$ with $\Gamma_{5678}\theta_{\pm}=\pm\theta_{\pm}$ so that, keeping into account the
representation we chose for $\Gamma_{11}$, (\ref{gam11}), $\theta_{\pm}$ has the following
decomposition in terms of 16-component spinors
\begin{equation}
\theta_\pm=\vecto{\vartheta^1_{\pm}}{\vartheta^2_\pm} \, ,
\end{equation}
The gauge conditions that should be imposed to fix $\kappa$-symmetry are different on $\theta_+$ and
on $\theta_-$~\cite{Astolfi:2009qh} and read
\begin{equation}
\label{kappacond} \Gamma_{-} \theta_- =0~~~\spa~~~\Gamma_{4956}\theta_+=\theta_+
\end{equation}
It is thus useful to rotate the $\theta_+$ spinor so as to impose also on the rotated spinor the same
gauge condition we have on $\theta_-$. This is done by defining $\widetilde\theta_+$ according to
\begin{equation} \label{tildetheta} \theta_+=(I-\Gamma_{0456})\widetilde\theta_+
\end{equation}
Again we have the decomposition in terms of spinors of opposite chirality
\begin{equation}
\widetilde\theta_+=\vecto{\widetilde\vartheta^1_{+}}{\widetilde\vartheta^2_+} \, ,
\end{equation}
The gauge choice on $\widetilde\theta_+$ is thus $\Gamma_{-} \widetilde\theta_+ =0$.

It is then useful to define also a rotated 16-component spinor
$\hat\vartheta^2_+=\hat\gamma_4\widetilde\vartheta^2_+$ so that both $\widetilde\vartheta^{1}_+$ and
$\hat\vartheta^2_+$ have the same eigenvalue +1 of $\hat\gamma_9$. This rotations make the
quantization on this type IIA background very similar to that of the type IIB.

We can now define the rescaled 8-component spinors
\begin{equation}
\label{defS2}\widetilde\vartheta^{1}_+ = \frac{1}{\sqrt{c}}\vecto{S^1_+}{0}~~\spa~~~\hat\vartheta^2_+
= \frac{1}{\sqrt{c}}\vecto{S^2_+}{0} \, ,
\end{equation}
In the main text we used then the 8-component complex spinors
\begin{equation}
\psi=S^1_++i S^2_+~~~\spa~~~\psi^*=S^1_+-i S^2_+
\end{equation}
Let us now turn to $\theta_-$. Again to have the same eigenvalue +1 of $\hat\gamma_9$ for the upper
and the lower 16-component spinors,  we perform a rotation of $\vartheta_-^2$ with $\hat\gamma_4$
according to $\hat\vartheta_-^2=\gamma_4\vartheta_-^2$. We can now define as before the rescaled
8-component spinors
\begin{equation}
\label{defS3}\widetilde\vartheta^{1}_- = \frac{1}{\sqrt{c}}\vecto{S^1_-}{0}~~\spa~~~\hat\vartheta^2_-
= \frac{1}{\sqrt{c}}\vecto{S^2_-}{0} \, ,
\end{equation}
In the main text we then used then the 8-component complex spinors
\begin{equation}
\chi=S^1_-+i S^2_-~~~\spa~~~\chi^*=S^1_--i S^2_-
\end{equation}

\end{appendix}



\providecommand{\href}[2]{#2}\begingroup\raggedright\endgroup

\end{document}